# Weedy Adaptation in *Setaria* spp.: VII. Seed Germination Heteroblasty in *Setaria faberi*

Milton J. Haar and Jack Dekker

Weed Biology Laboratory, Department of Agronomy, Iowa State University
Ames, Iowa 50011, USA

**Summary**

The dormancy status of *S. faberi* seed at abscission was assessed with reference to tiller and panicle development. Seed from a single genetic line were grown under field, greenhouse and controlled environment growth chamber conditions. At abscission, a small fraction (<10%) of *S. faberi* seed germinated under favorable conditions. Seed were dissected and germination of caryopses and embryos also tested. Removal of seed structures exterior to the embryo increased percentage germination. As the seed rain progressed mean percentage germination and variation among samples increased. Changes in germination were correlated with tiller development and relative time of seed maturity within a panicle. Seed produced on tillers that developed earlier were more likely to be dormant than seed from later-developing tillers. Seed that matured later on a panicle were more likely to germinate than seed that matured earlier on the same panicle. A consistent trend toward later maturing seed having less dormancy was found for seed grown under different environments which implies an inherent or parental source for variation in giant foxtail seed dormancy. The variation in percentage germination at abscission and following stratification treatments indicates that the *S. faberi* seed rain consists of individual seeds, possibly each with a different degree of dormancy.

## INTRODUCTION

*Setaria faberi* (giant foxtail) is among the worst weeds in U.S. agriculture (Knake, 1977). The success of *S. faberi* as a weed is attributable to many factors, including seed dormancy-germinability, dispersal over time. The biology of *S. faberi* seed developmental arrest and germination is poorly understood. Several studies have examined *S. faberi* dormancy-germination with a variety of treatments including: cool-dark-moist after-ripening (stratification; Stanway, 1971; Van den Born, 1971), heat (Tayorson and Brown, 1977), alternating temperatures (Fausey and Renner, 1997), seed scarification (Morre and Fletchall, 1963), hull puncture (Morre and Fletchall, 1963; Stanway, 1971), water stress (Blackshaw et al. 1981; Taylorson, 1986; Manthey and Nalewaja, 1987) and plant hormones (Kollman and Staniforth 1972). These treatments did not result in complete germination. A lack of consistency in methods of collection, storage and testing have made previous studies difficult to compare and interpret. Much of what is reported of *S. faberi* seed germinability is in apparent conflict. This may be a consequence of the many factors that influence germinability, and of an incomplete description of the history of the seed used in those investigations. The seed of *S. faberi* consists of a hull (tightly joined lemma and palea) that encloses the caryopsis (embryo and endosperm surrounded by the caryopsis coat) (Narayanaswami, 1956; Rost, 1973, 1975). *S. faberi* seed is shed from the parent plant primarily in a nongerminable, but viable, state of developmental arrest. Studies have concluded that all seed shed from the parent plant is dormant in *S. pumila* (Povalitis, 1956; Peters and Yokum, 1961; Rost, 1975) and *S. verticillata* (Lee, 1979), "almost completely" dormant in *S. pumila* (Taylorson, 1966; Kollman, 1970) and *S. faberi* (Stanway, 1971), or shed with variable germinability (*S. glauca*: Norris and Schoner, 1980; *S. viridis*: Martin, 1943). Viviparous (precocious) germination in *S. viridis* populations was observed in 1896 (Hubbard, 1915). Variable germinability may be a function of the role played by seed tissues and structures with time and development. Variation in seed germinability may also be a function of the foxtail species and genotype, the parental growth environment, and the seed germination environment. Research has indicated that germinability may be influenced by different structures within the foxtail seed. Differential germinability of mature seed was influenced by the hull (Peters and Yokum, 1961; Nieto-Hatem, 1963; Biswas et aI., 1970; Stanway, 1971; Rost, 1975), the caryopsis (Nieto-Hatem, 1963; Rost, 1975) and the isolated embryo (Rost, 1975). A previous study (Dekker et al., 1996) provided evidence that *S. faberi* seed germinability is based on the actions of the embryo, caryopsis, and hull compartments. These studies provided evidence for a dynamic, developmental model of giant foxtail germinability regulation resulting in phenotypes with a wide range of germinability shed from an individual panicle. These diverse germinability phenotypes were found at all stages of development, but particularly when the seed is shed and the soil seed bank is replenished.

Within a group of *S. faberi* seeds, a single level of dormancy does not exist, but that diversity in the requirements for germination among seeds from a panicle is the rule. In an effort to better understand seed dormancy in *S. faberi* we investigated the degree and variation of seed dormancy-germination with reference to plant development and environment. We hypothesized that the parent plant influences seed dormancy in *S. faberi* and that changes in seed dormancy could be correlated with plant architecture (tillering) and development (relative time of seed abscission and location within a panicle). Anthesis and abscission provided easily identifiable markers of plant and seed development. Variation in seed dormancy also results from differences in environment during seed development. Separation of environmental and developmental influences on seed dormancy is difficult, but can be overcome by experiments in controlled environmental conditions and collection of seed in a frequent and consistent manner. This study examined dormancy for seed from a single genotype of *S. faberi* grown under controlled, greenhouse and field conditions. We hypothesized that changing environmental conditions (field, greenhouse) during seed development would affect the degree and variation of giant foxtail seed dormancy, and that in a non-variable environment (growth chambers) an endogenous level of variation in seed dormancy induction would also be observed.

## MATERIALS AND METHODS

**Plant Material**.

*S. faberi* populations exhibit low genotypic variability relative to other weedy foxtail species (Wang, Wendel, and Dekker, 1995a, b). Despite this, and to ensure that the results obtained were not artifacts of genotypic variation, a single seed (single-seed descent propagation scheme; seed lot number 1816; Dekker et al., 1996) was used to propagate the seed used in these studies. A single seed provided a single plant, which was vegetatively separated several times, and the resulting plantlets grown in a glasshouse. These plants were transplanted into a field nursery near Ames, Iowa in the growing season of 1991. When panicles emerged from these plants, individual inflorescences were covered with pollination bags to prevent the introduction of foreign pollen. The seed were harvested from these panicles and stored in controlled environmental conditions (45-50% relative air humidity, dark, 4°-6°C) in the long-term seed storage facility in the Agronomy Building, Iowa State University, Ames, Iowa, USA until used in these investigations.

**Growth Conditions**

**Growth cabinets**. Controlled environment experiments were conducted in a growth chamber (SG30, Hoffman Man. Co., Albany, OR) set at 16:8 h light:dark and constant temperature (26°C). Radiation was provided with 15 (Phillips) high output fluorescent and 12 incandescent bulbs each 52W set 1.25 m above the growth chamber floor. Radiation intensity was 450±50 $\mu moles/m^{-2}/s^{-1}$ at approximate primary panicle level, 40 cm from floor of growth chamber. Seeds germinated in a 1:2:2 (v:v) mixture of soil, peat moss and perlite (Silbrico Corp., Hodgkin, IL, USA) under growth chamber conditions. Ten days after sowing, seedlings were transplanted to 15 cm diameter plastic pots filled with the same soil mixture. Thirty plants were used per experiment. Plants were given tap water as needed, approximately every other day, and fertilized every 2 wk with 30 mg N per plant.

**Greenhouse**. *S. faberi* plants grown in the greenhouse were started in soil, same mixture as above, transferred 10 d after planting to 10 cm diameter clay pots then at the five to six leaf stage, transplanted into 20 cm diameter clay pots. Fifteen ml Osmocote fertilizer (Grace-Sierra Horticultural Products, Milpitas, CA) was applied at second transplant and again 60 d later. Plants were watered (tap) as needed approximately every other day. Greenhouse studies took place in winter-spring. Seeds were planted for the greenhouse experiment (Ames, Iowa) on January 10, flowering began March 16, most seeds abscissed during April-June. Only natural lighting was used.

Field. Seasonal field studies took place in summer-autumn. Plants for the field experiment were started in the greenhouse and transplanted to the field (Ames, Iowa) at the five to six leaf stage (June 1), and most seeds abscissed during August-September. No additional water or fertilizer was applied.

**Seed collection**

At the first sign of anthesis, panicles were covered with a plastic mesh bag (7.6 x 25.4 cm Delnet bags, Applied Extrusion Inc., Middletown, DE) and secured with a wire twist tie. The bags collected seed as it abscised and prevented cross pollination. Recently abscised seed was harvested every 3 to 5 d by gently shaking the panicle then removing the plastic mesh bag. Harvests occurred from the time shattering began until seed was completely shed from a panicle. The *S. faberi* seed rain in the field typically occurs from late July until killing frost (first week of October, average for central Iowa).

In the controlled environment experiments, panicles were designated primary, secondary or tertiary based on tillering pattern (figure 1). Each plant had a single primary (1°) panicle at the end of the main culm. Branches that formed at the nodes of the main culm bore secondary (2°) panicles and branches that arose at the nodes of secondary branches bore tertiary (3°) panicles. Tiller development was a sequential process with a consistent pattern and duration.

**Figure 1**. Schematic description of *Setaria faberi* seed cohorts from primary (1°), secondary (2°) and tertiary (3°) panicle types: calendar date (time) of seed abscission (CD); seed position

on panicle at time of abscission (PP, PDP); cohort of seed from entire panicle based on age (PA, time of first flower of panicle).

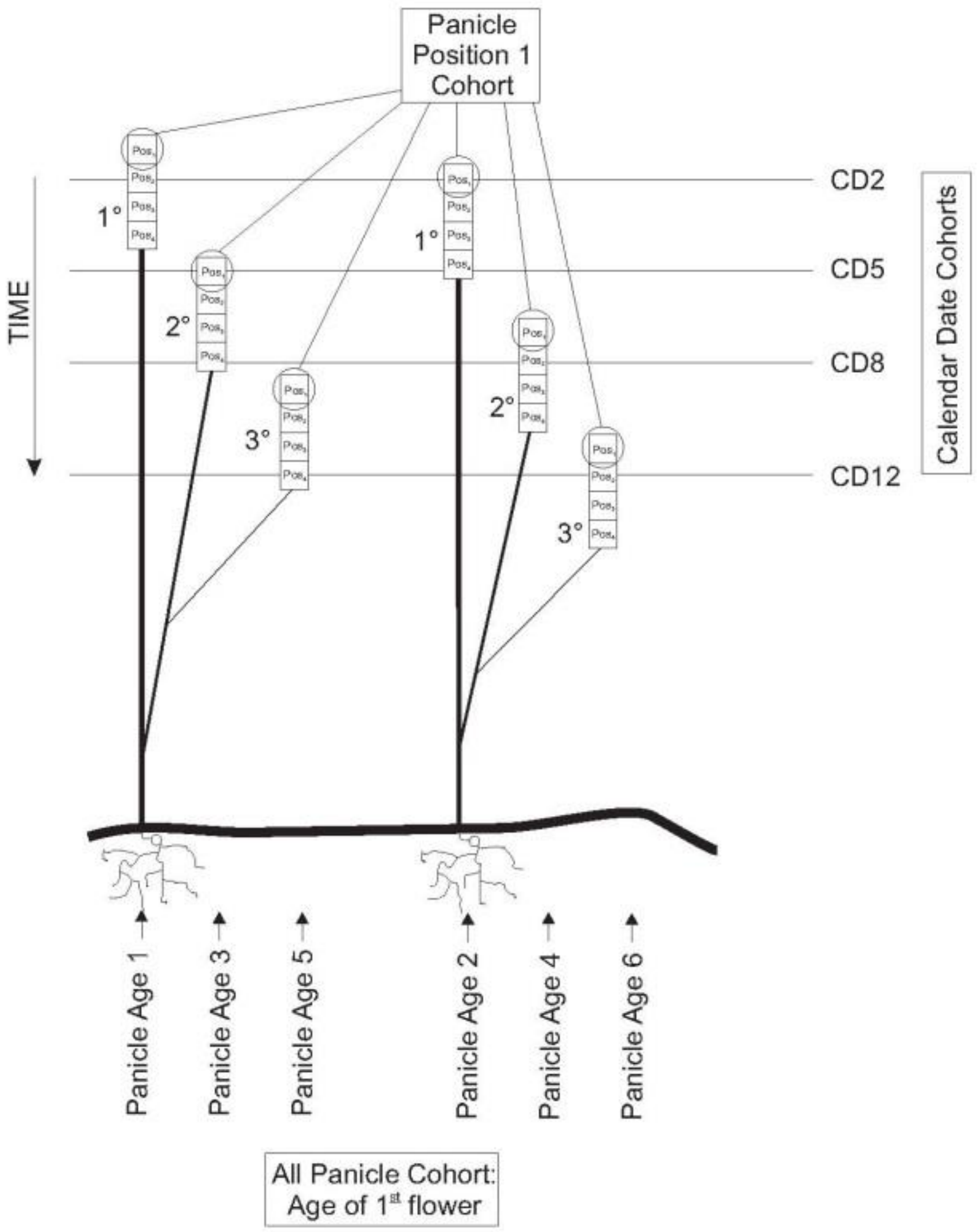

**Germination assays**

Germination tests were conducted immediately following harvest. Only dark brown mature seeds were used. For controlled environment experiments, germination of isolated caryopses and embryos was also tested (Dekker et al., 1996). Seeds were dissected under a microscope by using forceps and a scalpel. To facilitate embryo excision, caryopses were allowed to imbibe for 2 to 4 h under germination conditions. Care was taken to remove the caryopsis coat from over isolated embryos. Ten to 20 seeds or their dissected components were then put on moist (2 ml distilled water) blue blotter germination paper (5 cm) in 5-cm-diameter glass petri dishes sealed against water loss with parafilm and placed in a germination cabinet at 26°C and constant light, PPFD 640±40 µmoles/$m^{-2}/s^{-1}$. Percentage germination was recorded after 10 d. The criterion for seed germination was emergence of the coleorhiza or coleoptile. Evidence of embryo growth qualified as germination for the dissected components (Dekker et al., 1996).

**Stratification after-ripening**

Seed not tested for germination at harvest were stratified (cool-dark-moist) in petri dishes with moist sand (1 ml distilled water per 6 g sand), double wrapped in aluminum foil (dark) and kept at 4°C for 2, 4, or 8 wk). Following stratification, a sample was removed from sand under green light and germination tested as described above. Seed not tested for a treatment was left in petri dish, rewrapped in foil and returned to stratification conditions.

**Analysis**

Analysis of variation (ANOVA) was performed for all experiments, and the mean percentage germination was calculated for all treatments. Seeds, caryopses and embryos were grouped into cohorts for analysis in several ways to determine the influence of plant architecture (1°, 2° or 3° tiller panicles), calendar date of seed abscission (CD; regardless of tiller, panicle or position on panicle), position of the seed as it developed on an individual panicle (PP, PDP; calculated by the number of days after the first anthesis event on an individual panicle, regardless of tiller or calendar date), or age of the panicle (PA; dated from the day the first flower, anthesis, occurred on an individual panicle, regardless of position or calendar date) (figure 1).

## RESULTS

### Seed, Caryopis and Embryo Germination

**Germination at abscission**. At abscission, germination of field- and greenhouse-grown seed was 8.9% and 4.9%, respectively (Table 1). No seed germination at abscission was observed in any controlled environment experiment: the presence of surrounding caryopsis and hull tissues inhibited embryo germination. Removal of the hull increased germination. Caryopsis germination was low (about 1%), but greater than seed germination at that time. Isolated embryos had the greatest germination at abscission, between 20.1% and 40.5%. The naked embryo itself can be dormant as indicated by the many isolated embryos that did not initially germinate.

**Table 1**. Germination (%) and number (n) of *S. faberi* seeds, caryopses and embryos in response to stratification (weeks at 4° C, dark, moist). Seed developed under different types of environmental conditions: growth chamber (GC 1, GC 2, GC 3), field or greenhouse. [1] Means within columns for a compartment followed by the same letter are not different at p = 0.05 according to paired t-tests.

| Seed Compartment | Weeks Stratification | Germination (%) | | | | |
|---|---|---|---|---|---|---|
| | | Environment | | | | |
| | | GC1[1] | GC2 | GC3 | Field | Greenhouse |
| | | n = 77 | n = 275 | n = 406 | n = 298 | n = 448 |
| seed | 0 | 0.0 a | 0.0 a | 0.0 a | 8.9 a | 4.9 a |
| | 2 | 0.2 a | 0.0 a | 0.0 a | 6.9 a | 7.1 b |
| | 4 | - | 1.7 a | 0.0 a | 26.8 b | 32.3 c |
| | 8 | 5.8 b | 64.8 b | 13.9 b | 44.5 c | 77.2 d |
| | | | | | | |
| caryopsis | | n = 78 | n = 277 | n = 409 | | |
| | 0 | 0.5 a | 0.3 a | 1.2 a | | |
| | 2 | 5.1 a | 14.3 b | 3.9 a | | |
| | 4 | - | 49.2 c | 9.5 b | | |
| | 8 | 65.8 b | 78.2 d | 20.7 c | | |
| | | | | | | |
| embryo | | n = 81 | n = 370 | n = 430 | | |
| | 0 | 39.8 a | 20.1 a | 40.5 a | | |
| | 2 | 80.1 b | 83.4 b | 76.3 b | | |
| | 4 | - | 96.9 c | 94.2 c | | |
| | 8 | 98.3 c | 98.9 c | 94.9 c | | |

**Effect of cold-dark-moist stratification**. Stratification after-ripening (4° C, dark, moist) treatments increased seed, caryopsis and embryo germination in all experimental environments (Table 1). The magnitude of response varied by seed compartment, environment and experiment. Seed from the greenhouse was the most responsive to stratification followed by those from the field, and then the growth chambers. Two or four week stratification treatments increased germination of seed grown in field or greenhouse, but not in growth chamber experiments. Eight week stratification increased germination in all experiments, with the greatest in the greenhouse. The same stratification duration resulted in considerable variation in seed germination in different environments (e.g., 8 weeks: GC1, 5.8% versus greenhouse, 77.2%). Caryopses germination increased more than seeds, and less than embryos, with the same stratification treatments within an experiment. Caryopsis germination increased with each additional stratification period, but the amount of increase varied considerably between experimental runs (e.g., 8 weeks: GC2, 78.2% versus GC3, 20.7%). Embryo percentage germination increased more than that of seeds and caryopses with similar stratification within each controlled environment experiment. After four weeks of stratification, isolated embryos were essentially all germinable.

**Panicle tillers**. For comparisons in which a difference in germination occurred among seed or their dissected components, samples from later-developing panicles had the higher percentage germination. In controlled environment conditions, tiller development was a sequential process with a consistent pattern and duration (Figure 2); which was highly correlated with CD and PP cohorts (r= 0.90 to 0.96). Tiller panicle development began with the emergence of the primary (10) panicle at the terminal end of the main culm; followed by branches forming at the nodes of the main culm, these tillers were referred to as secondary (2°). Tertiary (3°) tillers aroseat nodes of the secondary tillers. Development of secondary and tertiary tillers began with lower nodes.

**Figure 2**. Time (days after first anthesis on a plant) of *Setaria faberi* flowering (□, ○; first to last anthesis) and seed rain (■, ●; first to last abscission, harvest) for primary (1°), secondary (2°) and tertiary (3°) panicle types from two growth chambers experiments GC1 (□, ■) and GC2 (○, ●)

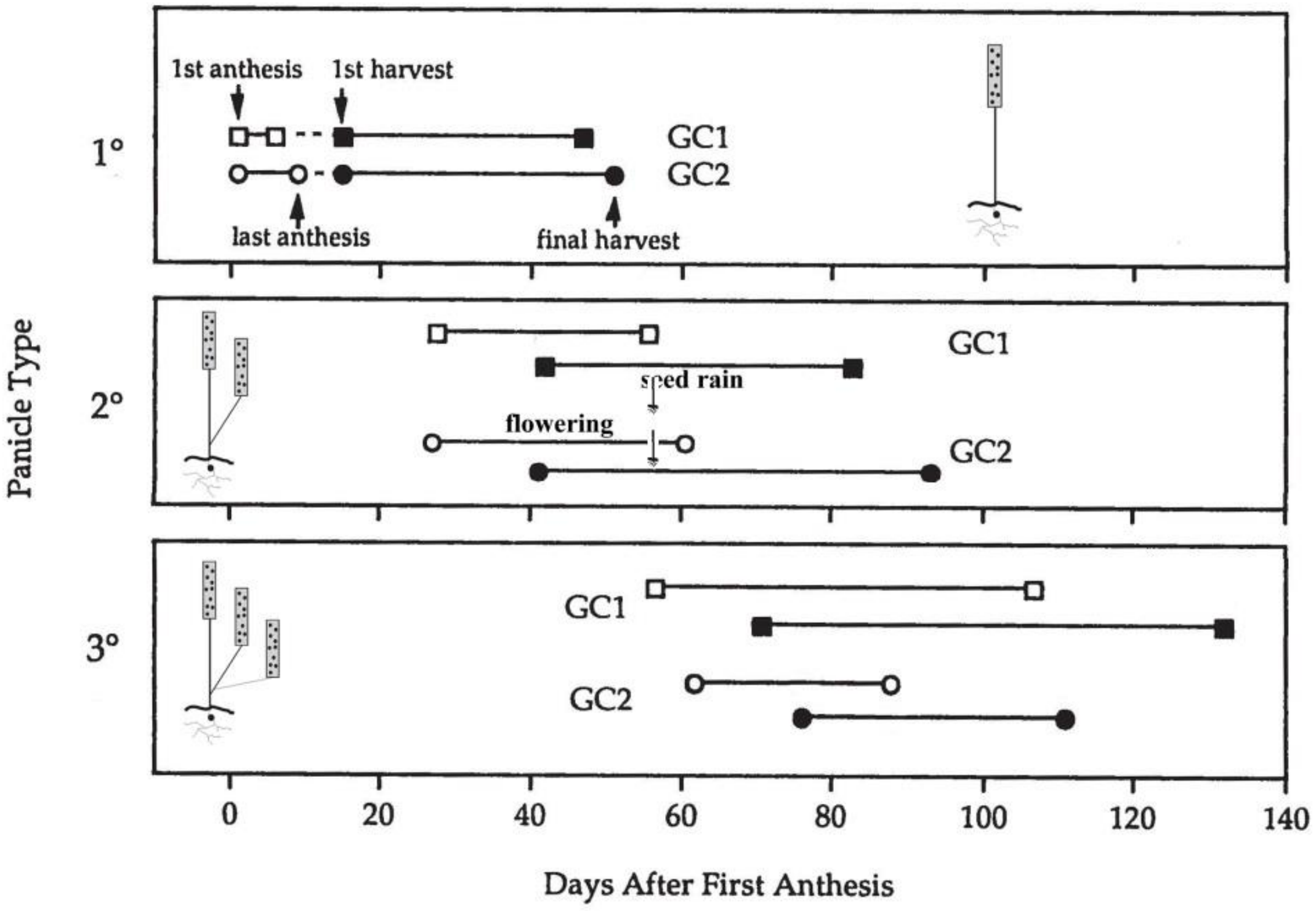

No differences were observed in the germination of seed or caryopses among panicle types in the second growth chamber experiment (GC2; table 2). In the third controlled environment experiment (GC3), germination in seeds from 1° panicles was less than that from 2° or 3° panicles after eight weeks of stratification. In GC3, caryopsis germination was less in 1° panicles than that in 2° (at abscission), and in 3° (4 and 8 wks stratification). Embryos from 2° panicles had higher germination than in 1° panicles in GC3 at abscission and after 2 and 4 wks stratification. Embryos from 3° panicles had higher germination than from 1° panicles in GC3 (at abscission; after 4 weeks stratification) and GC2 (2 wks stratification). Following eight weeks of stratification, embryo germination from all panicles types was above 90%.

**Table 2**. Germination (%) and number (n) of seeds, caryopses and embryos from primary (1°), secondary (2°), or tertiary (3°) *S. faberi* panicles in response to stratification (weeks at 4° C, dark, moist) in two different growth chamber experiments (GC2, GC3). [1] Means within columns for a compartment and experiment followed by the same letter are not different at p = 0.05 according to paired t-tests.

| Seed Compartment | Experiment | Panicle Type | Germination % – Weeks of Stratification: 0 | 2 | 4 | 8 |
|---|---|---|---|---|---|---|
| seed | GC2 | | n = 109 | n = 52 | n = 70 | n = 44 |
| | | 1° | 0.0 a | 0.0 a | 0.0 a | - |
| | | 2° | 0.0 a | 0.0 a | 3.2 a | 61.5 a |
| | | 3° | 0.0 a | 0.0 a | 1.4 a | 70.2 a |
| | GC3 | | n = 158 | n = 65 | n = 112 | n = 71 |
| | | 1° | 0.0 a | 0.0 a | 0.0 a | 5.0 a |
| | | 2° | 0.0 a | 0.0 a | 0.4 a | 19.2 b |
| | | 3° | 0.0 a | 0.0 a | 0.0 a | 23.7 b |
| caryopsis | GC2 | | n = 111 | n = 51 | n = 72 | n = 43 |
| | | 1° | 1.1 a | 12.9 a | 48.7 a | - |
| | | 2° | 0.0 a | 10.5 a | 42.0 a | 78.5 a |
| | | 3° | 0.0 a | 17.6 a | 56.0 a | 77.6 a |
| | GC3 | | n = 160 | n = 66 | n = 111 | n = 72 |
| | | 1° | 0.1 a | 2.1 a | 6.3 a | 14.4 a |
| | | 2° | 2.1 b | 5.0 a | 10.0 ab | 24.0 ab |
| | | 3° | 1.3 ab | 7.1 a | 17.1 b | 29.4 b |
| embryo | GC2 | | n = 152 | n = 67 | n = 105 | n = 46 |
| | | 1° | 16.7 a | 69.2 a | 94.5 a | - |
| | | 2° | 18.9 a | 84.2 ab | 96.1 a | 100 a |
| | | 3° | 22.9 a | 90.9 b | 98.8 a | 99.7 a |
| | GC3 | | n = 181 | n = 66 | n = 111 | n = 72 |
| | | 1° | 28.5 a | 64.0 a | 87.1 a | 95.5 a |
| | | 2° | 47.9 b | 92.5 b | 99.6 b | 95.0 a |
| | | 3° | 43.2 b | 65.7 a | 100 b | 92.7 a |

**Seed Cohort Germination**

**Panicle position seed cohorts**. Germination differed among cohorts of seeds with similar positions on the panicle axis (PP) that were grown in both the field ($p < 0.05$) and greenhouse ($p < 0.0001$) (table 3). No differences in germination were observed among panicle position seed or caryopsis cohorts in any controlled environment experiment (GC1-3; $p > 0.05$), but there were differences among PP embryo cohorts in GC3 ($p < 0.05$). Germination percent, and variation within sample times, increased with developmental time along the panicle axis in PDP cohorts from field (seed; figure 3, top), greenhouse (seed; figure 3, middle) and growth chamber (embryo; figure 3, bottom) environments in which differences were observed.

**Table 3**. Analysis of variance (ANOVA) for parameters describing biological and environmental influences during development on seed, caryopsis and embryo germination (%). Data are from five experiments: three growth chamber (GC1-3), field and greenhouse conditions. Calendar date (CD) cohort refers to seed that matured on the same date. Panicle development position (PDP) cohort is the relative location within a panicle where a seed developed. Panicle age (PA) cohort determined by time (date) of first anthesis on a panicle. [1] Degree of significance: NS, $P > 0.05$; *, $P < 0.05$; **, $P < 0.01$; ***, $P < 0.001$; ****, $P < 0.0001$.

| Source of Variation | Seed Compartment | GC1 | GC2 | GC3 | Field | Greenhouse |
|---|---|---|---|---|---|---|
| Stratification | seed | **[1] | **** | **** | **** | **** |
| | caryopsis | **** | **** | **** | | |
| | embryo | *** | **** | **** | | |
| | | | | | | |
| Calendar Date | seed | NS | NS | NS | * | **** |
| | caryopsis | NS | NS | NS | | |
| | embryo | NS | *** | NS | | |
| | | | | | | |
| Panicle Position | seed | NS | NS | NS | * | **** |
| | caryopsis | NS | NS | NS | | |
| | embryo | NS | NS | * | | |
| | | | | | | |
| Panicle Age | seed | NS | NS | NS | * | **** |
| | caryopsis | NS | ** | NS | | |
| | embryo | NS | ** | *** | | |

**Figure 3**. Relationship between germination (%) and seed position on panicle at time of abscission (PDP, panicle developmental position) cohorts from plants grown in the field (seed; top), greenhouse (seed; middle) and third growth chamber study (GC3; embryo; bottom). A PDP cohort consists of seed that abscissed the same number of days after first anthesis on a panicle. Bars represent standard errors of the mean and are not shown when less than marker size.

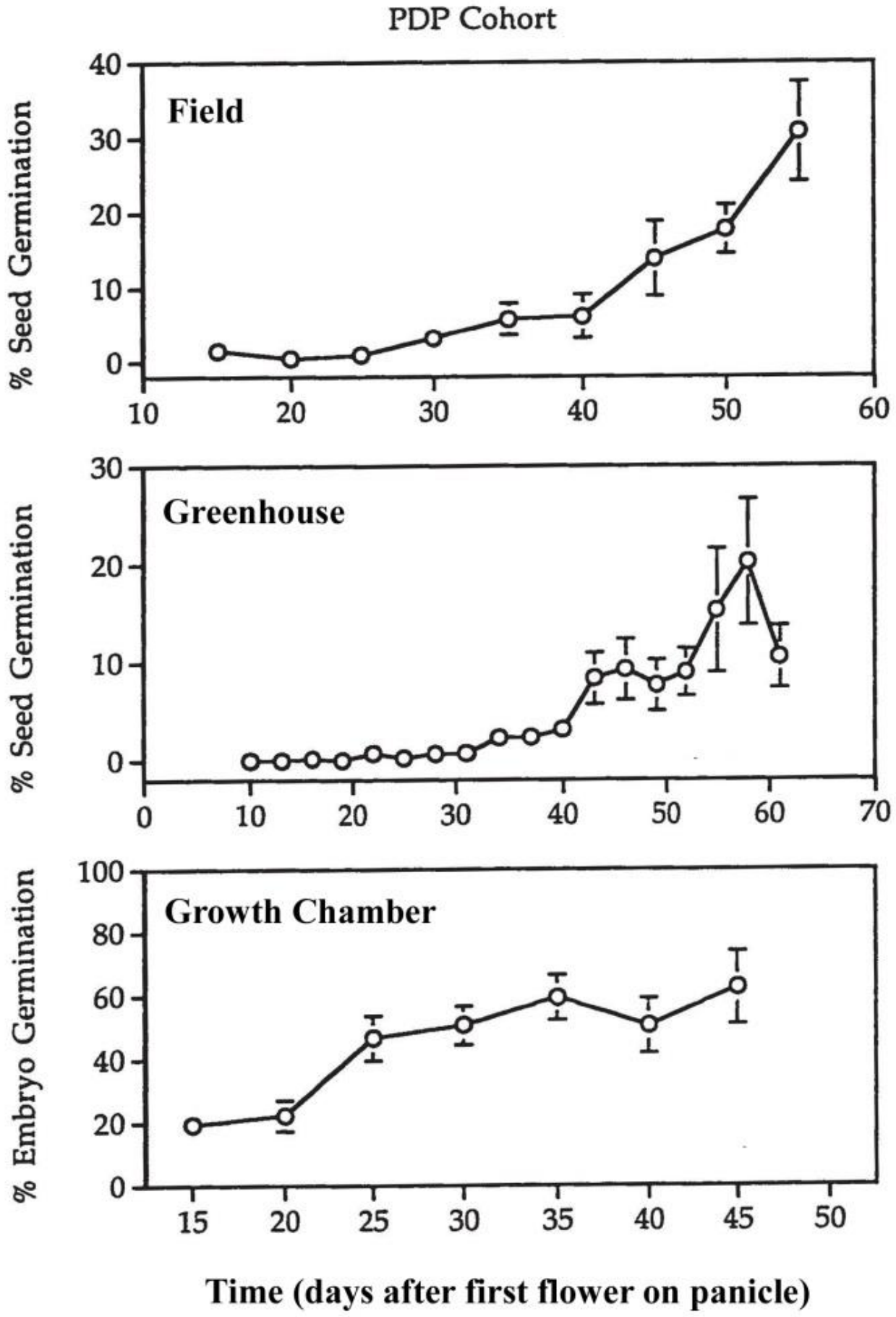

**Calendar date of seed abscission cohorts**. Germination differed among cohorts with different calendar dates of seed abscission (CD) from *S. faberi* plants grown in the field (seed; $p < 0.05$) and greenhouse (seed; $p < 0.0001$) (table 3). No differences in seed or caryopsis germination were found among CD cohorts for controlled environment experiments (GC1-3; $p > 0.05$), but there were differences among CD embryo cohorts in GC2 ($p < 0.001$). Germination, and variation within sample dates, increased within CD cohorts with calendar time from field (seed; figure 4, top) and greenhouse (seed; figure 4, middle) environments as the season progressed. These observations were similar to those in panicle position cohorts. No seasonal pattern was observed for embryo germination in the growth chamber (GC2) in GC2, although the highest embryo germination occurred at a later abscission time (d 92; figure 4, bottom).

**Panicle age cohorts**. Seed germination differed among panicle cohorts of different ages (PA; based on date of first panicle flower) *S. faberi* plants grown in the field ($p < 0.05$) and greenhouse ($p < 0.0001$), but not in those from growth chamber experiments ($p > 0.05$) (table 3). Differences in germination among PA cohorts were observed in caryopses (GC2, $p < 0.01$) and embryos (GC2, $p < 0.01$; GC3, $p < 0.001$) from the growth chamber. No consistent pattern of changes in seed, caryopsis and embryo germination was observed among PA cohorts (figure 5). Most germination differences were obscured by the variability resulting from grouping all seeds from a panicle together in PA cohorts, regardless of position on the panicle, calendar date of abscission or tiller. Despite this, embryo germination in the later (youngest) developing panicles in the growth chamber (GC3) was greater than that observed in the first (oldest) panicles (figure 5, bottom). Additionally, seed germination in the later (youngest) panicles was greater than those in some of the first (oldest) panicles grown in the greenhouse (figure 5, middle top), and field (figure 5, top).

**Figure 4**. Relationship between germination (%) and seed cohorts with the same calendar date (CD) of abscission from the field (seed; top), greenhouse (seed; middle) and second growth chamber study (GC2; embryo; bottom). A CD cohort consists of recently abscissed seeds harvested on the same date. Bars represent standard errors of the mean and are not shown when less than marker size.

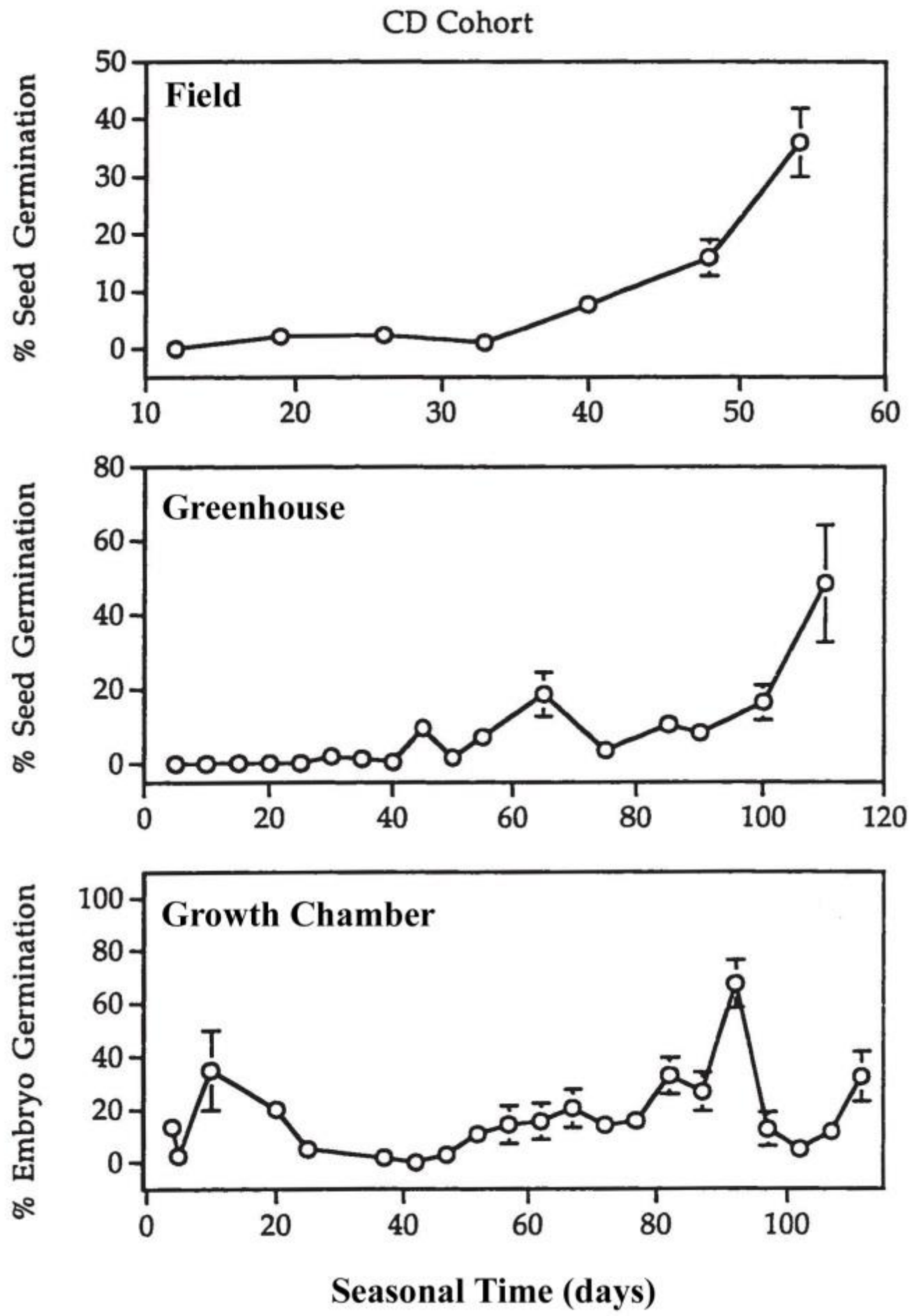

## DISCUSSION

Germination tests at abscission, stratification treatments and seed dissection were used to evaluate seed dormancy (Table 4). Stratification was effective in promoting germination of giant foxtail seed or their dissected components in each experiment, and provided important information about dormancy and germinability states not revealed in germination assays conducted at abscission. Comparison of isolated embryo germination with caryopses or whole seed germination indicated that tissues surrounding the embryo inhibit its germination. This is consistent with previous studies indicating whole giant foxtail seed dormancy is controlled by its several component parts (Dekker et aI., 1996).

**Figure 5**. Relationship between germination (%) and cohorts of seed at abscission taken from the entire panicle based on age (PA; time of first flower on the panicle) from field (seed; top), greenhouse (seed; middle top), second growth chamber (GC2; caryopsis germination after 4 weeks stratification; middle bottom), and third growth chamber (GC3; embryo germination after 4 weeks stratification; bottom) experiments. Bars represent standard errors of the mean and are not shown when less than marker size.

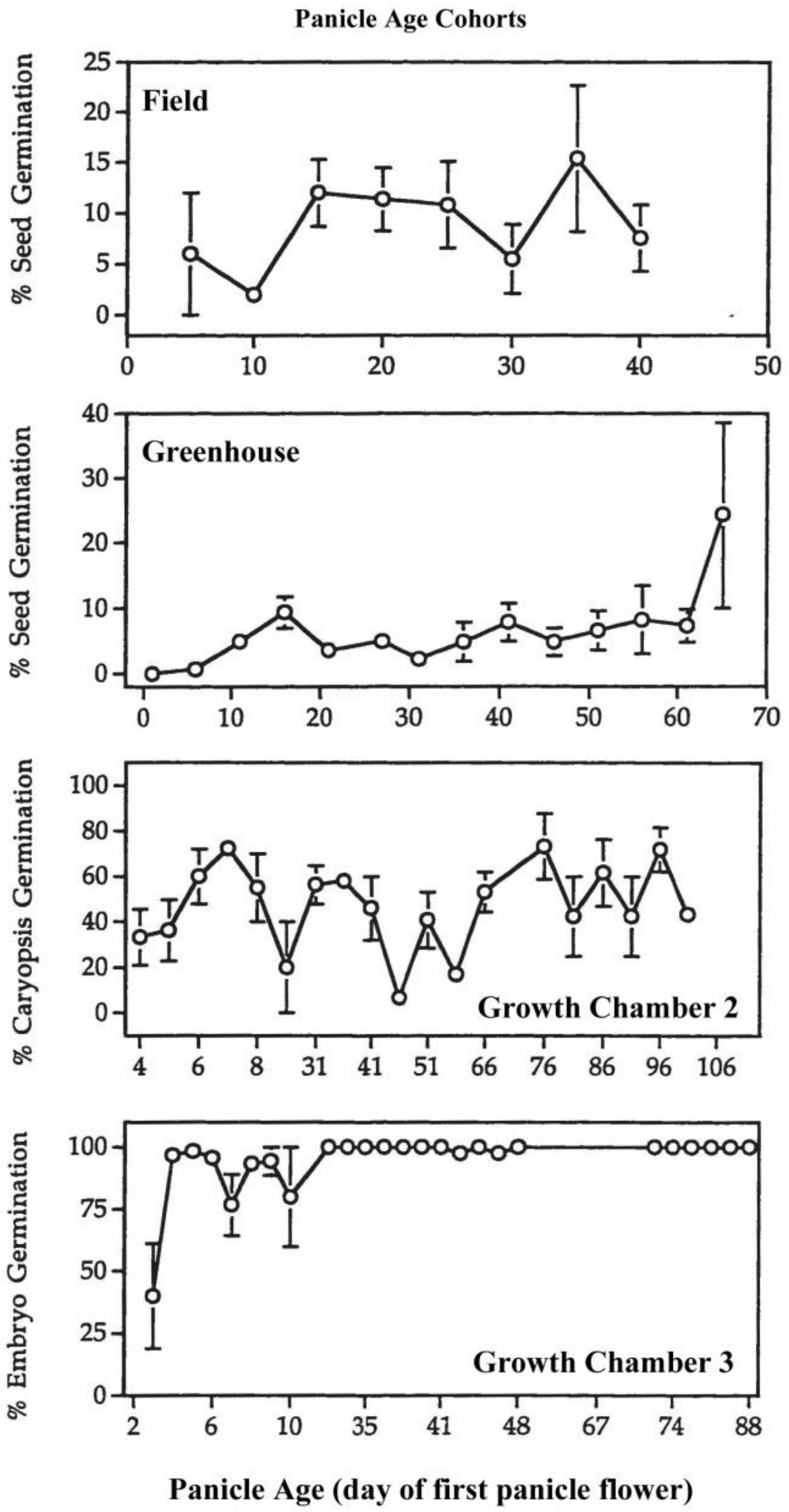

**Table 4**. Summary of seed germination results.

| Plant-Panicle-Seed | | Environment | | | | |
|---|---|---|---|---|---|---|
| | | GC1 | GC2 | GC3 | GH | Field |
| **Plant** (Table 1) | | **$GERM_{ABSC}$ → $GERM_{8week}$** | | | | |
| | embryo | 40 → 98% | 20 → 99% | 41 → 95% | | |
| | caryopsis | 1 → 66% | 1 → 78% | 1 → 21% | | |
| | seed | 0 → 6% | 0 → 65% | 0 →14% | 5 → 77% | 9 → 45% |
| | | | | | | |
| **Plant** (Figure 4) | | **Seasonal Time** | | | | |
| | embryo | | high germ late | | | |
| | seed | | | | germ increases with time | |
| | | | | | | |
| **Panicle** (Figure 3) | | **Seed Position on Panicle** | | | | |
| | embryo | | | embryo germ increases down panicle axis with time | | |
| | seed | | | | embryo germ increases down panicle axis with time | |
| **Panicle** (Figure 3) | | **Panicle Age** | | | | |
| | embryo | | | embryo germ in first-oldest panicle < youngest | | |
| | seed | | | | oldest two panicle cohorts germ < youngest two cohorts | 2nd oldest panicle cohort germ < several older cohorts |
| | | | | | | |
| **Plant** (Table 2) | | | **$GERM_{ABSC}$ → $GERM_{8week}$** | | | |
| **1° Panicle** | embryo | | $GERM_{ABSC}$ = 17% | 29% → 96% | | |
| | caryopsis | | $GERM_{ABSC}$ = 1% | 1% → 14% | | |
| | seed | | $GERM_{ABSC}$ = 0% | 0% → 5% | | |
| **2° Panicle** | embryo | | 19% → 100% | 48% → 95% | | |
| | caryopsis | | 0% → 79% | 2% → 24% | | |
| | seed | | 0% → 62% | 0% → 20% | | |
| **3° Panicle** | embryo | | 23% → 100% | 43% → 93% | | |
| | caryopsis | | 0% → 78% | 1% → 29% | | |
| | seed | | 0% → 70% | 0% → 24% | | |
| **1° vs. 2° vs. 3°** | embryo | | 1° < 3° | 1° < 2°<br>1° < 3° | | |
| | caryopsis | | | 1° < 2°<br>1° < 3° | | |
| | seed | | 1° < 2°<br>1° < 3° | | | |
| | | **GC1** | **GC2** | **GC3** | **GH** | **Field** |

**Dormancy Induction and Environment**

Giant foxtail seed grown in different environments have the potential to vary considerably in their germination. Results support the hypothesis that seed grown in different environments would vary in degree of dormancy: large differences in germination were found among the experimental regimens. No seed germinated at harvest from plants grown in growth chambers, while low numbers of seed germinated at harvest from field- and greenhouse-grown plants. Stratification revealed greater germination (less dormancy) in greenhouse-grown plants compared to those grown in the field. Giant foxtail seed grown under apparently equivalent conditions also varied in percentage germination. Stratification revealed a high degree of latent variation in germination not apparent at harvest in growth chamber grown seed. Over tenfold differences in seed germination were found among the several growth chamber experiments (e.g. Table 1, eight week stratification). Four to five-fold differences in caryopsis germination between growth chamber experiments not apparent at harvest were also revealed by stratification (e.g. Table 1, 4,8 week stratification). Finally, two-fold differences in germination between embryos from different growth chamber experiments were observed at harvest. These results suggest that the dormancy induction mechanism(s) that operate during embryogenesis and seed development (caryopsis and hull dormancy) are sensitive to environmental conditions. This sensitivity may be an inherent developmental or genetic trait that acts either independently of, or in conjunction with individual spikelet microenvironment. Support for endogenous, developmental control of dormancy induction is provided by comparing the large differences in seed, caryopsis and embryo dormancy among the growth cabinet experiments (Table 1) with the consistent panicle development sequence for the plants that produced those seeds (Figure 2). These results are also very compelling considering the large numbers of seed (1604 petri dish tests), caryopses (764 petri dish tests) and embryos (881 petri dish tests) evaluated to reach these conclusions (Table 1).

**Tiller Development and Dormancy**

The tiller type on which a giant foxtail seed develops is an important source of seed, caryopsis and embryo dormancy heterogeneity. Under controlled environmental conditions, subtle differences in germination among seed from different panicle tiller types was revealed. In general, seed from panicles on earlier emerging (primary) tillers was more dormant than seed from later-emerging secondary and tertiary tiller panicles (Table 2). Germination of seed from primary panicles was less than germination of seed from secondary and tertiary tillers (e.g., Table I, GC3: eight week stratification). Germination of caryopses from primary panicles was less than that from secondary (e.g., GC3: at harvest) or tertiary (e.g., GC3 four, eight week stratification). Embryos from primary panicles were more dormant compared to those from secondary (e.g., GC3: at harvest; two, four week stratification) or tertiary panicles (e.g. GC2: two week stratification; GC3: at harvest, four week stratification). Calendar date cohorts provided an indirect indication of the influence tiller type has on heterogeneous seed dormancy phenotypes. As the seed rain season progressed, seed and embryo germination and variability increased (Figure 3; Table 5). This decrease in dormancy with time was probably not due to changes in the environment, as it occurred under controlled growth chamber conditions. A more likely explanation is that increased germination was correlated with plant and panicle development from the more dormant primary to the less dormant tertiary panicles. The time of development (PP cohorts) provided a less sensitive indication of the influence tiller panicle type has on germination variation at abscission. Although PP cohorts were highly correlated with tiller development, sensitivity to differences in germination associated with PP cohort was less because these cohort groups bulk seed from all parts of the panicle in a single mean. No consistent pattern of change in germination with PP cohorts was observed, but when differences were found, the later-developing seed and embryos were less dormant than those maturing earlier (Figure 4b, 5b).

**Seed Position in the Panicle and Dormancy**

The relative time of development (PDP), or panicle position, of a seed within an individual panicle is another important influence of seed dormancy in giant foxtail. Seed developing earlier in an individual panicle were more dormant that those maturing later in the same panicle. The later maturing seed were also more variable in germination than earlier, more dormant seed. This contribution of development time within a panicle to germination variability occurred in giant foxtail plants grown in the field, greenhouse and growth chamber. The effect of panicle position had a much larger influence on seed germination than did tiller type.

**Cumulative Influences of Development, Morphology and Environment on Dormancy**

Seed shed from giant foxtail plants during the seed rain of one season are mostly dormant at abscission, with only a small fraction capable of germination immediately under favorable conditions. As the seed rain season progresses the percentage germination, as well as the variability in germination requirements of those seeds, increases. The sources of seed dormancy heterogeneity are both environmental and biological. The more variable environmental conditions resulted in greater seed heterogeneity. Biological sources of germination variability are associated with morphology and development. The panicle type, relative position within the panicle, and the interaction of seed parts with differing amounts dormancy, are all factors which contribute to a heterogeneous seed rain.

**Weedy Adaptation and Seed Dormancy Heterogeneity**

Observations of seed dormancy provided here are not adequately explained by concepts of seed dormancy as a single state with a "trigger" mechanism that releases dormant seed to a second state, germination. The variation in dormancy observed at abscission, and the differences in response to stratification, indicate that giant foxtail produces individual seeds, each with a different germinability state (Trevewas, 1987). These dormancy states arise during development as a function of each seed's structural and physiological components as modified by parental influences and environment (Come 1980/81; Dekker et al., 1996; Silver town, 1984). The production of seed with a variety of germination requirements is an advantage to giant foxtail. Dormancy heterogeneity among seed in the soil seed bank permits some germination to occur over an extended period of time or under a wide range of environmental conditions. This increases the likelihood of some giant foxtail plants avoiding unfavorable environmental conditions to produce seed.

## REFFERENCES

Banting, J. D., E. S. Molberg and J. Gebhardt. 1973. Seasonal emergence and persistence of green foxtail. Can. J. Plant Sci. 53:369-376.

Biniak, B. M., and R J. Aldrich. 1986. Reducing velvetleaf *(Abutilon theophrasti)* and giant foxtail *(Setaria faberi*) seed production with simulated-roller herbicide applications. Weed Sci. 34:256-259.

Biswas, P. K., A. G. Chakrabarti, H. A. Collins and R B. Bettis. 1970. Histochemical studies of weed seed dormancy. Weed Sci. 18:106-109.

Blackshaw, R. E., E. H. Stobbe, C. F. Shaykewich and W. Woodbury. 1981. Influence of soil temperature and soil moisture on green foxtail *(Setaria viridis)* establishment in wheat *(Triticum aestivum).* Weed Sci. 29:179-185.

Cavers, P. B. 1983. Seed demography. Can. J. Bot. 61:3578-3590.

Chavarria, P. L. 1986. Seed dormancy characteristics in six weed species as affected by after-ripening temperature and field conditions. Ph.D. dissertation. Iowa State University, Ames.

Coleman, J. S., K. D. M. McConnaughay and D. D. Ackerly. 1994. Interpreting phenotypic variation in plants. Trends in Ecology and Evolution. 9:187-190.

Come, D. 1980/81. Problems of embryonal dormancy as exemplified by apple embryo. J. Bot. 29:145-156.

Dekker, I., B. Dekker, H. Hilhorst, and C. Karssen. 1996. Weedy adaptation in *Setaria spp.*: IV. Changes in the germinative capacity of *S. faberi* (Poaceae) embryos with development from anthesis to after abscission. Am. J. Bot. 83:979-991.

Dyer, W. E. 1995. Exploiting weed seed dormancy and germination requirements through agronomic practices. Weed Sci. 43:498-503.

Fenner, M. 1991. The effects of the parent environment on seed germinability. Seed Sci. Res. 1:75-84

Grime, J. P. 1981. The role of seed dormancy in vegetation dynamics. Ann. of AppI. BioI. 98:555-558.

Gutterman, Y. 1992. Maternal effects on seeds during development. Chpt. 2 in Seeds: The ecology of regeneration in plant communities. ed. M. Fenner. C. A. B. International. Wallingford, U. K. pp. 27-59.

Haar, M. J. and J. Dekker, 1998. Weedy adaptation in *Setaria spp.* V: Seed production in giant *(Setaria faberi)*, green *(S. viridis)* and yellow *(S. glauca)* foxtail. [Chapter 2]

Harper, J. H. 1977. Population Biology of Plants. Academic Press Ltd. San Diego, CA 92101. 892 pp.

Jain, S. K. 1982. Variation and adaptive role of seed dormancy in some annual grassland species. Bot. Gaz. 143:101-106.

King, L. J. 1952. Germination and chemical control of the giant foxtail grass. Contrib. Boyce Thompson Inst. 16:469-487.

Knake, E. L. 1977. Giant foxtail: The most serious annual grass weed in the Midwest. Weeds Today 9:19-20.

Kollman, G. E. 1970. Germination-dormancy and promoter-inhibitor relationships in *Setaria lutescens* seeds. Ph. D. dissertation, Iowa State University, Ames.

Lee, S. M., and P. B. Cavers. 1981. The effects of shade on growth, development, and resource allocation patterns of three species of foxtail *(Setaria).* Can. J. Bot. 59:1776-1785.

Li, H. W., C. J. Meng and T. M. Liu. 1935. Problems in the breeding of millet *(Setaria italica* (L.) Beauv.). J. Am. Soc. Agron. 27:963-970.

Manthey, D. R. and J. D. Nalewaja. 1987. Germination of two foxtail *(Setaria)* species. Weed Tech. 1:302-304.

Martin, J. N. 1943. Germination studies of the seeds of some common weeds. Iowa Acad. Sci. 50:221-228.

Morre, D. J. and O. H. Fletchall. 1963. Germination-regulating mechanism of giant foxtail *(Setaria faberi).* Univ. of Missouri College of Agriculture Experiment Station Research Bulletin 829. pp. 1-25.

Narayanswami, S. 1956. Structure and development of the caryopsis in some Indian millets. VI. *Setaria italica.* Bot. Gaz. 118:112-122.

Nieto-Hatem, J. 1963. Seed dormancy in *Setaria lutescens.* Ph.D. dissertation, Iowa State University, Ames, lA.

Norris, R. F. and C. A. Schoner, Jr. 1980. Yellow foxtail *(Setaria iutescens)* biotype studies: Dormancy and gennination. Weed Sci. 28:159-163.

Peters, R. A. and H. C. Yokum. 1961. Progress report on a study of the germination and growth of yellow foxtail *(Setaria glauca* (L.) Beauv.) Proc. Northeast Weed Control Conf. 15:350-355.

Povilaitis, B. 1956. Dormancy studies with seeds of various weed species. Proc. Int. Seed Test. Assoc. 21:87-111.

Rost, R. L. 1971. Structural and histochemical investigation of dormant and nondormant caryopses of *Setaria lutescens* (Gramineae). Ph.D. dissertation, Iowa State University, Ames, IA.

Rost, R. L. 1975. The morphology of germination in *Setaria lutescens* (Gramineae): The effects of covering structures and chemical inhibitors on dormant and non-dormant florets. Ann. Bot. 39:21-30.

[SAS] Statistical Analysis Systems. 1989. SAS/STATS User's Guide, version 6, 4th ed., vol. 2. SAS Institute, Inc. Cary, NC. 846 pp.

Scheiner, S. M. 1993. Genetics and evolution of phenotypic plasticity. Annu. Rev. Ecol. Syst. 24:35-68.

Schoner, C. *A.*, R. F. Norris and W. Chilcote. 1978. Yellow foxtail *(Setaria lutescens)* biotype studies: Growth and morphological characteristics. Weed Sci. 26:632-636.

Schreiber, M. M. and L. R. Oliver. 1971. Two new varieties of *Setaria viridis.* Weed Sci. 19:424-427.

Sells, G. D. 1965. C02:02 ratios in relation to weed seed germination. Ph.D. dissertation, Iowa State University, Ames, IA.

Silvertown, J. W. 1984. Phenotypic variety in seed germination behavior: The ontogeny and evolution of somatic polymorphism in seeds. Am. Nat. 124:1-16.

Simpson, G. M. 1990. Seed dormancy in grasses. Cambridge University Press. New York, NY. 297 pp.

Stanway, V. 1971. Laboratory germination of giant foxtail *(Setaria faberiJ* Herrm. at different stages of germination. Proc. Assoc. Off. Seed Anal. 61:85-90.

Sultan, S. E. 1987. Evolutionary implications of phenotypic plasticity in plants. Evol. BioI. 21:127-178.

Taylorson, R. B. 1986. Water-stress induced germination of giant foxtail *(Setaria faberi)* seeds. Weed Sci. 34:871-875.

Taylorson, R. B. 1987. Environmental and chemical manipulation of weed seed dormancy. Rev. Weed Sci. 3:135-154.

Till-Boutraud, 1., X. Rebound, P. Brabant, M. Lefarn, B. Rherissi, F. Vedel, H. Darmency. 1992. Outcrossing and hybridization in wild and cultivated foxtail millets: Consequences for the release of transgenic crops. Theoretical and Applied Genetics. 83:940-946.

Trevewas, A. J. 1987. Timing and memory processes in seed embryo dormancy: A conceptual paradigm for plant development questions. Bioessays 6:87-92.

Van den Born, W. H. 1971. Green foxtail: Seed dormancy, germination and growth. Can. J. Plant Sci. 51:53-59.

Wang, R.,J. F. Wendel andJ. H. Dekker. 1995a. Weedy adaptation in *Setaria spp.* 1. Isozyme analysis of genetic diversity and population genetic structure in S. *viridis.* Am. J. Bot. 82:308-317.

Wang, R., J. Wendel and J. Dekker. 1995b. Weedy adaptation in *Setaria spp.* II. Genetic diversity and population genetic structure in S. *glauca*, S. *geniculate* and S. *faberi* (Poaceae). Am. J. Bot. 82:1031-39.